\documentclass[conference,10pt]{IEEEtran}
\RequirePackage{snapshot}
\pdfoutput=1
\makeatletter
\def\ps@headings{%
\def\@oddhead{\mbox{}\scriptsize\rightmark \hfil \thepage}%
\def\@evenhead{\scriptsize\thepage \hfil \leftmark\mbox{}}%
\def\@oddfoot{}%
\def\@evenfoot{}}
\makeatother
\usepackage{epsfig,url,times,amsmath}
\usepackage{subfigure}
\usepackage{pifont}
\usepackage{color}    
\usepackage{xspace}
\usepackage{multirow}
\usepackage{cite}
\usepackage[font=small,labelfont=bf]{caption}

\newcommand{\pr}[2]{\ensuremath{\mathsf{P}(#1\,|\,#2)}}
\newcommand{\prior}[1]{\ensuremath{\mathsf{P}(#1)}}

\begin{document}
\title{Mixture Models of Endhost Network Traffic}

\author{\IEEEauthorblockN{John Mark Agosta\IEEEauthorrefmark{1}, Jaideep Chandrashekar\IEEEauthorrefmark{2}, Mark Crovella\IEEEauthorrefmark{3}, Nina Taft\IEEEauthorrefmark{2} and Daniel Ting\IEEEauthorrefmark{4} }
\IEEEauthorblockA{\IEEEauthorrefmark{1}Toyota ITC}
\IEEEauthorblockA{\IEEEauthorrefmark{2}Technicolor Research}
\IEEEauthorblockA{\IEEEauthorrefmark{3}Boston University}
\IEEEauthorblockA{\IEEEauthorrefmark{4}Facebook} }

\maketitle

\begin{abstract}
In this work
we focus on modeling a little studied type of traffic, namely the network
traffic generated from endhosts.  We introduce a parsimonious parametric
model of the marginal distribution for connection arrivals. We employ mixture
models based on a convex combination of component distributions with both heavy
and light-tails. These models can be fitted with high accuracy using maximum
likelihood techniques. Our methodology assumes that the underlying user data can be fitted to one of
many modeling options,  and we apply Bayesian model selection criteria as a
rigorous way to choose the preferred combination of components. Our experiments
show that a simple Pareto-exponential mixture model is preferred for a wide
range of users, over both simpler and more complex alternatives. This model has
the desirable property of modeling the entire distribution, effectively
segmenting the traffic into the heavy-tailed as well as the non-heavy-tailed
components. We illustrate that this technique has the flexibility to capture the
wide diversity of user behaviors.

\end{abstract}

\section{Introduction}
\label{sec:intro}

In the last decade or so there has been a tremendous amount of research
done in the area of Internet traffic modeling (e.g.,
\cite{
  paxson94:poisson_failure_model,
  crovella97:www_self_similarity,
  Leland1993selfsimilar} to name a few).
Traffic models are helpful in solving a
wide range of problems, including traffic engineering, service
provisioning, routing, and network performance evaluation.  To date,
however, the vast majority of traffic modeling research has focused on
traffic seen inside a network: at routers, gateways, or servers.
Relatively little work has been done to model traffic as seen at
endhosts, such as laptops or desktops.

The paucity of endhost traffic models is limiting, because many problems
can benefit from an understanding of the nature of endhost traffic. Recently there is increased interest in modeling and describing
the behavior of enterprise end users \cite{coffeeshop, user-classify}.
IT management is driving this trend, as it
faces an increasingly heterogeneous computing environment. Autonomic
computing is heading towards self-diagnosis for fault identification,
and endhost profiles are being explored for security purposes
\cite{wren09} and resource management. For example, in
\cite{host-control-sigcomm08} the authors design mechanisms to allow
hosts to participate in network management, traffic engineering and
other operational decisions by explicitly controlling host traffic. To
better calibrate such applications, a deep understanding of end user
traffic is needed.


The most likely reason that endhost traffic models are so scarce is that
it is difficult to obtain the raw measurements needed, since those
measurements require the express consent of each user in a sufficiently
large set.   Furthermore, such measurements essentially require installing a
collection tool directly on each user's machine --- a tool whose management
requires considerable goodwill from the affected users.

The value of endhost models combined with the difficulty of endhost
instrumentation have motivated some efforts that have tried
to infer endhost traffic properties from an observation point inside the
network \cite{enduser-inference,user-classify}.  While such approaches
have shed some light, they are fundamentally limited --- for example,
when users are mobile.  What is needed for a comprehensive view of
endhost traffic is a measurement tool that moves with the user and
continues to observe network traffic as the user switches between
different networks and different environments (e.g., work and home).

In this work we deploy such a tool and analyze its outputs to develop
models for end user traffic.  We study a population of 270
enterprise users over a period of five weeks (\S\ref{sec:dataset}).  Our tool collects all
packet headers entering and exiting the machine, on all networking
interfaces.  To accomplish this, we solicited enterprise employees to
sign up on a voluntary basis for the trace collection.  Participants
explicitly gave consent for data collection; each user downloaded and
installed the data collection software on their personal machines.

Starting with this rich dataset, we focus on careful
  modeling of user activity, in particular the arrival process of flow
  initiations.   A main focus of our work is developing a robust method
  for distributional modeling of flow arrivals.   To that end, we go
  beyond simple parameter estimation and attack the  \emph{model
    selection} problem.

In our approach we overcome concerns about the use of \emph{goodness-of-fit
testing} for choosing probability models, and in particular for power law models,
about estimating their {\em scaling parameter}.   Commonly used
  methods for estimating the scaling parameter include the \emph{Hill
    estimator} and least-squares regression on a log-log plot of the
  histogram.  The Hill estimator is notoriously tricky since it relies on estimating a cut-off
below which the central part of the distribution is disregarded, to concentrate on just
the scaling parameter of the small subset of data in the tail
\cite{
Newman.2005,
willinger-topology}, while
extracting the scaling parameter by performing a
least-squares regression has been shown
to little more than a misleading heuristic, of questionable statistical value.
In \cite{Clauset:2007}, the authors highlight the lack of care
pervasive in the literature on power laws, and apply a rigorous approach to
applying goodness of fit methods.   In the process they review numerous
power law claims that have been made, and find that
claims of power law tails among well-known supposedly-``power law''
datasets are not supported by the data.   In our work,
we demonstrate an efficient estimator that uses the entire data set
(rather than just the tail).

Hence, our first contribution is in
modeling endhost traffic using \emph{mixture models} (\S\ref{sec:mix-models})
to estimate model parameters.
A mixture model
is a convex combination of component distributions, where the
parameters of the component distributions as well as the mixture
parameter are estimated from data.


To discriminate among the class of mixture models we need a criterion,
the commonly applied one being goodness-of-fit.
The limitation of this approach is that goodness-of-fit
tests, and their associated $P$-values, are meant to rule out hypotheses
(i.e. to reject the hypothesis). This is certainly useful for steering
data collection, but they do not provide an acceptance criterion.
The best one can hope for with this method, in a statistical sense, is
to say that a given model has not been ruled out by the data.
In our situation, with effectively an endless stream of data as a source,
in the limit of large data, the
probability of any model decreases along with goodness of fit's $P$-values, and
the result is that any reasonable model will be rejected.
 What is needed to confirm the proper choice of model is not
model fitting but rather model selection, a different problem with a
different statistical basis.

Model selection methods give us a quantitative criterion that lets us explore
a wider class of models than has hitherto been considered.
 Thus we do not presuppose a
single parametric distribution model; instead we start with a class of {\em
 nested} mixture models (i.e. a family of models where one is a subset
of another) and use {\em Bayes Factors}~\cite{KassRaftery:1995} to select the
best model in the class for a user's data.
For the large sample
sizes that we consider, the log Bayes Factor can be well approximated by
the difference of two models' {\em Bayesian Information Criteria} (BIC).
Since it is a
requirement for Bayes Factors comparisons to compare both models on the full
sample data (not just the distributional tail), as a side benefit we produce complete
models as they vary over the set of users.


Our second contribution is thus to validate an approach that makes available
a richer, non-parametric (in the sense of the number of parameters is not
fixed before model selection) class of models for traffic modeling.
We use Maximum Likelihood methods (\S\ref{sec:mix-model-params}) for
parameter estimation and validate (\S\ref{sec:mix-model-params-validation}) the accuracy
of our parameter estimation technique on synthetic data created from
mixture models. Our success with this method in modeling endhost traffic
suggests that it might be fruitful to explore using this modeling
technique to other heavy-tailed datasets of network measurements.





Our third contribution lies in the results of extensive application of
this method on our endhost traffic data (\S\ref{sec:results}).
We observe that the distribution of flow arrival counts can be generally
characterized as monotonically declining, from a mode at zero.   Hence
we can eliminate a vast majority of possible component distributions
(such as, e.g., Gaussian or Poisson) and concentrate on mixtures of
various exponential and Pareto distributions.    Since mixtures of
exponentials in particular constitute a very flexible framework,
restricting to these two distributions does not severely limit our modeling
ability.
Thus our model selection process considers various combinations of the
Pareto (P)
 component with one or more exponential (E) distribution components to form
 a nested family of mixture models.
This family includes mixtures such as EP, EEP, and P.

We find that for the metrics we study (flow arrivals and idle
period lengths), the vast majority of users are well modeled by the EP
distribution; a much smaller number are better modeled by P and EEP
models.  Here the flexibility of our approach is a strength, because our
method does not insist that all users need to be described by the same
model.

Our final contribution lies in examining the results of our modeling.  We
expose and highlight strong invariants across users (\S\ref{sec:results}), but also illustrate
the nature of the diversity among different users.  For example, we
demonstrate that tail properties and mixing fractions differ
dramatically across our set of users.  Finally in \S\ref{sec:applns} we look into which applications and services contribute
traffic to the exponential and pareto components of our models.

\section{Related Work}
\label{sec:related}


Heavy tailed statistics have been documented in numerous phenomena in
network traffic; in the popularity of web pages
\cite{Breslau99webcaching}, in traffic demands
\cite{Feldmann01derivingtraffic}; in network topology
\cite{willinger-topology}, in TCP inter-arrival times
~\cite{feldmann-TCParrivals}, in wireless LAN traffic \cite{wireless},
and many others. As mentioned earlier, most of this work analyzes
traffic collected from inside the network at locations where anywhere from
hundreds to millions of users' traffic is aggregated. To the
best of our knowledge, our work is the first to study connection
traffic generated directly on user laptops.

The seminal work by Leland
et al.~\cite{ Leland1993selfsimilar} studied LAN traffic and
convincingly demonstrated that actual network traffic is self-similar or
long-range dependent in nature (i.e., bursty over a wide range of time
scales).  Our work differs in two ways.   First, that study's Ethernet
LAN data captured the aggregated traffic of many users, whereas we focus
on models for individual user traffic.   Second, we observe the power law
nature of traffic in the first-order statistics of traffic rates, rather
than in the second-order autocorrelation properties.  Both approaches
result in estimating a power-law parameter, but the meaning of the
parameters should not be confused.

Work closer to our study is reported in
\cite{CunhaBestavrosCrovella95,BBBC99}. Those studies captured HTTP
requests through instrumentation in web browsers or proxies, and so are
similar to ours in focusing on a traffic seen at a fixed set of endhosts.
Results from those studies were used in developing tools for generating
representative user-level HTTP traffic \cite{BarfordCrovella98}.
However, those studies did not look at the total set of an individual
endhost's pattern of connections over time.  This crucial
difference makes our results more useful for general traffic modeling. In fact,
as we shall see, the aggregate traffic on endhosts is influenced
strongly by applications other than the Web.  An important aspect of
this distinction is that our data also includes network traffic that is
machine generated (i.e. not user generated). Machine generated traffic
comes from background enterprise applications, chatty protocols, and the
many auto-update checking mechanisms (e.g., for software and firewall
rule updates) that are typically installed on corporate laptops.

Another end user study looks at data from Neti@Home users, and models
think time as well as bytes sent and received for TCP and UDP
connections \cite{neti-user-study}. They do not model traffic at the
granularity of connection arrivals.  In \cite{wren09}, the authors
report on the diversity in distributional tails of user behaviors. This
diversity is captured by a simple metric, the 99$^{th}$ percentile of
various distributions on user protocol traffic.   The models proposed
in our paper capture user tail diversity with richer
measures, such as the slope parameter $\alpha$ of the Pareto
distribution from the EP model.

The idea of using mixture models for Internet traffic has been proposed
in other contexts before \cite{FeldmannWhitt97}. That work proposes
using hyperexponential models to approximate heavy-tailed
distributions. Thus it is not about explicitly modeling data collected
from the Internet, but more about fundamental methods for approximations
of heavy-tailed distributions. The advantage of their work is that their
effort provides analytically tractable representations that can be used
subsequently for queueing theory models. However, the disadvantage (as
the authors acknowledge) is that their mixture models have a large
number of parameters. In our work, we obtain parsimonious models with a
small number of parameters.  All of our models range from having 1 to at
most 5 parameters; most users are well modeled using only 3
parameters.   Further, in constrast to the fitting-oriented approach
\cite{FeldmannWhitt97},  our work does
not assume the presence of a heavy-tailed component ahead of time.   It is
entirely possible for our mixture model to assign a negligible Pareto
component to a dataset.

\section{Dataset Description}
\label{sec:dataset}

The dataset used in this paper consists of traces collected at
270 enterprise end-hosts (90\% laptops), spanning a period of
approximately 5 weeks. Each end-host was associated with a unique user for the entire trace collection period, and ran a corporate standard build of Windows XP which included a number of enterprise IT applications.

Packet-level traces were collected {\em on} the end-hosts, rather than
at a network tap, providing a longitudinal view of the traffic even as the end-hosts moved in and out of the network and switched between interfaces (wired and wireless). The trace logging software included a wrapper around {\tt WinDump} to log packet headers and a homegrown application which sampled user-activity indicators (\# keystrokes, \# mouse clicks) and CPU load every second. The trace logging software tracked changes in IP address and network interface, and restarted the traffic trace collection at such times. The logged data was uploaded opportunistically a few times a day to a central server (the logging was paused during the upload). The trace collection effort yielded approximately 400 Gb of packet header traces. The packet traces were converted to flows (in the standard five-tuple sense) using BRO~\cite{bro}.

The starting time of each flow generates a point process in continuous  time that we bin over non-overlapping constant duration time-windows to create a time series for each user. Each user trace was binned for 8 different window sizes, starting at 4 seconds, and increasing in multiples of 2, up to 512 seconds. Each bin contains a count of the new flow arrivals. The {\em flow count} events within each time-window or {\em bin} are the random variables modeled in this work. In our datasets the median sample size was 9771 intervals, and 
the maximum was 264,000. Zeros could occur in bins because the host was turned off (or asleep), or else if the host was disconnected from the network during that bin. We filter out all such bins and in the resulting data, we see zeros only because there were no flows originated in that bin (and the machine was turned on). That being said, we mainly focus on modeling the flow events when the counts are nonzero since our goal is to characterize the flow traffic when the network is being used.

\section{Methodology}


\subsection{Mixture Models with Heavy Tails}
\label{sec:mix-models}

A \emph{mixture model} is a likelihood function composed of a convex
combination of probability densities. Such models are familiar in the
Statistics
literature,~\cite{EverittHand.1981}\cite{marinmengersenrobert:0204}
and have become a mainstay in the machine learning community ~\cite{JordanJacobs.1994}.    
A mixture model can be thought of as a hierarchical model where the mixing weights determine
the probability of each of the component models, which in turn generate the
sample points. Since all components share the same support, any sample point
could in principle have been generated by any component, but possibly with vanishingly
small probability.

A mixture model is defined by a probability density. For component densities, $f_i(x)$, and mixture fractions $m_i$,
the finite mixture model of $k$ components, with parameters $\mathbf{m}, \mathbf{\theta}$ is the convex combination given by:
\begin{eqnarray}
f(x\,|\,\mathbf{m}, \theta) = \sum^k_{i=1}{m_i}f_i(x\,|\,\theta_i),\\
{\hbox{s.t. }}\sum^k_{i=1}m_i = 1, m_i > 0\nonumber.
\end{eqnarray}
where the $\theta$ are the component parameters, and $\mathbf{m}
= m_1\ldots m_k$. The \emph{ degrees of freedom} of the model is the count of parameters, e.g. for $k$ components, each
with a single parameter, the full model will have $k + (k-1)$
parameters. 




We consider the following nested family of models: a Pareto only model labeled (P),
a mixture of one exponential and one Pareto (EP), and a mixture of two exponentials and one Pareto (EEP).
The ``pure power-law'' model we fit is
\begin{gather}
f(x\,|\,\alpha) = Cx^{-\alpha} = \frac{1}{\zeta(\alpha,x_{min})}x^{-\alpha}, x \in {\cal N} \tag{P}
\label{eqn:P}
\end{gather}
where $x$ takes on positive integer values, for which we use the discrete version of the Pareto density
(referred to also as the {\em Zeta}) in our models. The value of the Zeta function in the normalizing
constant for the discrete Pareto is
\begin{equation}
\label{eqn:zeta}
\zeta(\alpha,x_{min}) = \sum^{\infty}_{n=0}(n+ x_{min})^{-\alpha}.
\end{equation}

The exponential - Pareto model is defined as
\begin{gather}
f(x\,|\,\mathbf{m}, \lambda_1, \alpha)
=  m_1\lambda_1{e^{-\lambda_1{x}}} \tag{EP} \\
 {}+ (1-m_1)Cx^{-\alpha}\notag.
\label{eqn:EP}
\end{gather}
The mixture variable
adds another degree of freedom, revealing the relative contribution of the components.

The two exponential - Pareto mixture density model is:
\begin{gather}
f(x\,|\,\mathbf{m}, \lambda_1, \lambda_2, \alpha)
=  m_1\lambda_1{e^{-\lambda_1{x}}} \tag{EEP} \\
{}+ m_2\lambda_2{e^{-\lambda_2{x}}} \notag\\
{}+ (1-m_1-m_2)Cx^{-\alpha}\notag.
\label{eqn:EEP}
\end{gather}

We were motivated originally to consider these models
because  visual exploration of the data showed a traffic flow distributions with mode left-most, then a monotone decrease with a linear segment on a semi-log plot in the dense part of the distribution, followed by a long, heavy tail.


The intent behind using a family of models is to capture the diversity of each user's machine.
In principle, any combination of the 3 component distributions could be discovered, although in practice we always see a heavy-tailed component.
In terms of degrees of freedom, these are very parsimonious models; the EP model has 3 parameters, and the EEP has only 5.



\subsection{Estimating Model Parameters}
\label{sec:mix-model-params}

The model parameters are estimated using maximum likelihood.
The maximum likelihood estimate (MLE) has numerous attractive
qualities. If the model contains the true data generating distribution,
and is differentiable in quadratic mean (DQM) \cite{vaart},
the MLE converges to the
true parameters at a rate $O(1/\sqrt{n})$.
Pareto distributions and mixtures of DQM models satisfy
differentiability in quadratic mean.
Even if the model
does not contain the true data generating distribution, the MLE
converges to the best approximation to the true distribution within
the model's constraints at a rate $O(1/\sqrt{n})$. The MLE is also
 asymptotically efficient, so no other estimator can obtain
a better asymptotic variance than the MLE.

Instead of a conventional Expectation-Maximization (EM) methods,
we solved the MLE as constrained optimization problem
using an interior point method \cite{IntPoint}
to enforce the constraints on the model parameters. We found EM
converged slowly, probably due to the similar shapes of the components.
Interior point methods are iterative optimization methods that
enforce constraints by adding a weighted concave
 barrier function that steeply decreases to $-\infty$ at the boundary of the
constraint set, preventing the estimates from violating constraints.
The weight on the barrier
is decreased while using the previous solution
for initialization, and a new solution is computed.
The weight continues to be reduced
until the barrier becomes negligible.
A typical choice of barrier function
is $\log$.
Thus, for the EP model with log likelihood function $l$,
the constrained optimization problem
\begin{equation*}
\max_{\substack{m_1+m_2 = 1 \, m_1,m_2 > 0\\ \alpha > 1,\, \lambda > 0}} l( m_1,m_2,\alpha,\lambda; x)
\end{equation*}
may be solved by the
sequence of unconstrained problems
\begin{align*}
\max_{m_1,\alpha,\lambda} &\qquad  l(m_1,1-m_1,\alpha,\lambda; x) +
 c_1^{(t)} \log(\alpha-1) \\
& + c_2^{(t)} \log(\lambda) + c_3^{(t)} \log(1-m_1) + c_3^{(t)} \log(m_1)\\
\end{align*}
where $m_2$ has been replaced by $1-m_1$
and the weights on the barrier
$c_i^{(t)} \to 0^+$ as $t \to \infty$.
By convention, we take $log(x) = -\infty$
if $x \leq 0$.
These unconstrained problems can be solved using the {\tt optim()}
function in the statistical programming language R, which
implements a Quasi-Newton optimization method.
To exclude obviously bad solutions, we also added
constraints $\alpha < 4$ and
 $\lambda < 3.5$ so that the Pareto and
the exponential parameters did not grow not too large.


Since the mixture model typically contains local optima,
we performed the optimization multiple times
with random initializations to find the
global maximum. We also used small initial values of $c_i = 0.01$
for the regularization
parameters and reduced them to $c_i = 10^{-8}$ in 3 steps
to prevent the initial unconstrained problem and regularization path from
unduly influencing the search for the global maximum.

\subsection{Model Selection}
\label{sec:model-fits}

Given two probability models for the same sample, {\em model selection}
is a means of comparing which model is
more probable.
Our explanation of model selection, borrows extensively from Kass and Raftery~\cite{KassRaftery:1995}, and is based upon the numerical value of the comparative metric called a {\em Bayes Factor} ($BF$), .   Following Jefferies,
to quote, ``The Bayes Factor is a summary of
the evidence in favor on one scientific theory, represented by a statistical model, as opposed to another.''
This can be understood using the odds ratio form of Bayes rule, where the
posterior odds---the ratios of posteriors---between two models,
is expressed as the product of the $BF$ and the prior odds. So, for example
to compare the model $\mathcal{M}_P$ to the proposed
model $\mathcal{M}_{EP}$, the posterior odds will be
\begin{equation}
\label{eqn:posteriorOdds}
\frac{\pr{\mathcal{M}_{EP}}{\mathcal{D}}}{\pr{\mathcal{M}_P}{\mathcal{D}}}=%
\frac{\pr{\mathcal{D}}{\mathcal{M}_{EP}}}{\pr{\mathcal{D}}{\mathcal{M}_P}}%
\frac{\prior{\mathcal{M}_{EP}}}{\prior{\mathcal{M}_P}}
\end{equation}
where the middle term in this equation, the Bayes factor, $BF$, is defined as the ratio of
marginal likelihoods:
\begin{equation}
BF_{EP,P} = \frac{\pr{\mathcal{D}}{\mathcal{M}_{EP}}}{\pr{\mathcal{D}}{\mathcal{M}_P}}%
\end{equation}
The larger $BF$, the greater the weight of evidence for the EP model.
This criterion is similar to a maximum likelihood ratio, but
rather than taking the probability at the maximum, one integrates
over the range of parameters $\mathbf{\theta}$,
resulting in a correction for the degrees of freedom of the models.
Adding more parameters to a model and thus increasing its
degrees of freedom can only increase the likelihood at the maximum but does
not necessarily improve marginal likelihood. This criterion trades off simplicity with
accuracy---a built-in ``Occam's Razor."

\subsection{Interpreting The Weight of Evidence}

Interpreting the magnitude of a $BF$ is commonly
done by considering the ratio as an odds ratio, e.g., odds
of 20 to 1 in favor of the model in the numerator corresponds to
a $BF = 20$, or, using natural logs, $\log BF \simeq 3$.
Of course, the test is symmetric and the ratio may go either way. A negative $\log BF_{EP,P}$ is evidence {\em against} the EP model, in favor of P.
We give precedence to the conventional model, and hence require an log odds-ratio significantly greater than zero---we use 10---to chose EP.
If the EP model is selected, then we compute $\log BF_{EEP,EP}$. Again, if this factor is above 10, then EEP is selected, otherwise the final choice is EP.

Table~\ref{table:BF} shows a standard convention\cite{KassRaftery:1995} that we adopt for interpreting the strength of Bayes Factors with their suggested labels. Our threshold of 10 is well into the ``decisive'' range, corresponding to an odds ration of greater than 20,000. 
%

\begin{table}
\caption{Interpretation of Bayes Factor strengths}
\begin{tabular}{rccr}\hline
Odds   & $\log_{10}(BF)$  & $\log(BF)$    & Strength of comparison\\
\hline
20:1   & 1.3              &  3            &``substantial''\\
100:1  & 2                &  4.5          &``strong''\\
1000:1 & 3                &  7            &``decisive''\\
\hline
\end{tabular}
\label{table:BF}
\end{table}

\subsection{Approximation by BIC}

In practice the integral implied by $\pr{\mathcal{M}}{\mathcal{D}, \mathbf{\theta}}$
that requires a prior over the $\mathbf{\theta}$ is rarely done explicitly.
Experiments on our large sample data showed that likelihood values are infinitesimal and
strongly peaked around their maximum at $\mathbf{\hat{\theta}}$. Not surprisingly
numerical integration works poorly, so a common recourse is to approximate
the integral by the Laplace approximation.
\footnote{Also known in the literature as the \emph{ saddlepoint approximation}.~\cite{MacKay:2003}}
The Laplace approximation can be further approximated by the
Bayes Information Criterion (BIC). BIC is often presented as a
correction to maximum log likelihood to account for the degrees of
freedom of a model. The BIC is defined as
\begin{equation}
\mbox{BIC} =  \log\pr{\mathcal{D}}{\mathcal{M}, \mathbf{\hat{\theta}}} - \log(N)\cdot d/2
\end{equation}
where $N$ is the sample size and $d$ is the numbers of parameters in the model.
In our experimental work we computed both Laplace approximations and
BIC corrections and found to our satisfaction that they agreed with each other to within
a fraction of a percent on the data used.

With the BIC approximation, the log Bayes Factor becomes
\begin{equation}
\log BF_{EP,P} = \mbox{BIC}_{EP} - \mbox{BIC}_{P}
\label{eq:logBF}
\end{equation}
The BIC criterion is appealing as a standard procedure in that it can be applied even when
the priors on $\mathbf{\theta}$ are hard to choose.
\section{Validation}
\label{sec:mix-model-params-validation}

We validated our model-fitting and selection method in two ways, first to show that the estimates produced are accurate, and secondly that the selection mechanism we propose can distinguish among any of the three models. To do this we use synthetic data where the true value of the parameters of the generating data is known.  The test data consisted of pseudo-random samples with known parameters $\hat{\alpha},\hat{m_{\{1,2\}}},\hat{\lambda_{\{1,2\}}}$, generated from each of the three models in the family P, EP and EEP.  Since the test data is generated according to the same probability law to which the data is being fit, the models do not have to approximate the sample; we know that the data is in the same class as one of the three models. The same can be said of the Pareto-tail fitting procedure used, that is, by ignoring the dense part of the distribution contributed by the ``E'' component, will see a pure power-law sample.

\subsection{Estimation Accuracy}

The tail-fitting method used for comparison is a widely used~(e.g., \cite{papagiannaki04impact})
 tool~\cite{Crovella99estimatingthe} for
estimating the  $\alpha$ parameter of $\alpha$-stable distributions, based on a scaling property of sums of heavy-tailed random variables. An attractive property of this estimator is that it is nonparametric and easy to apply. Characteristic of Hill-estimator based methods~\cite{resnick.2007:heavy-tail} the method also estimates where the tail begins by computing a minimum value with which to select a range of the sample. We used a publicly available implementation, called {\bf aest}.

\begin{figure}
\includegraphics[width=\columnwidth,height=2.5in]{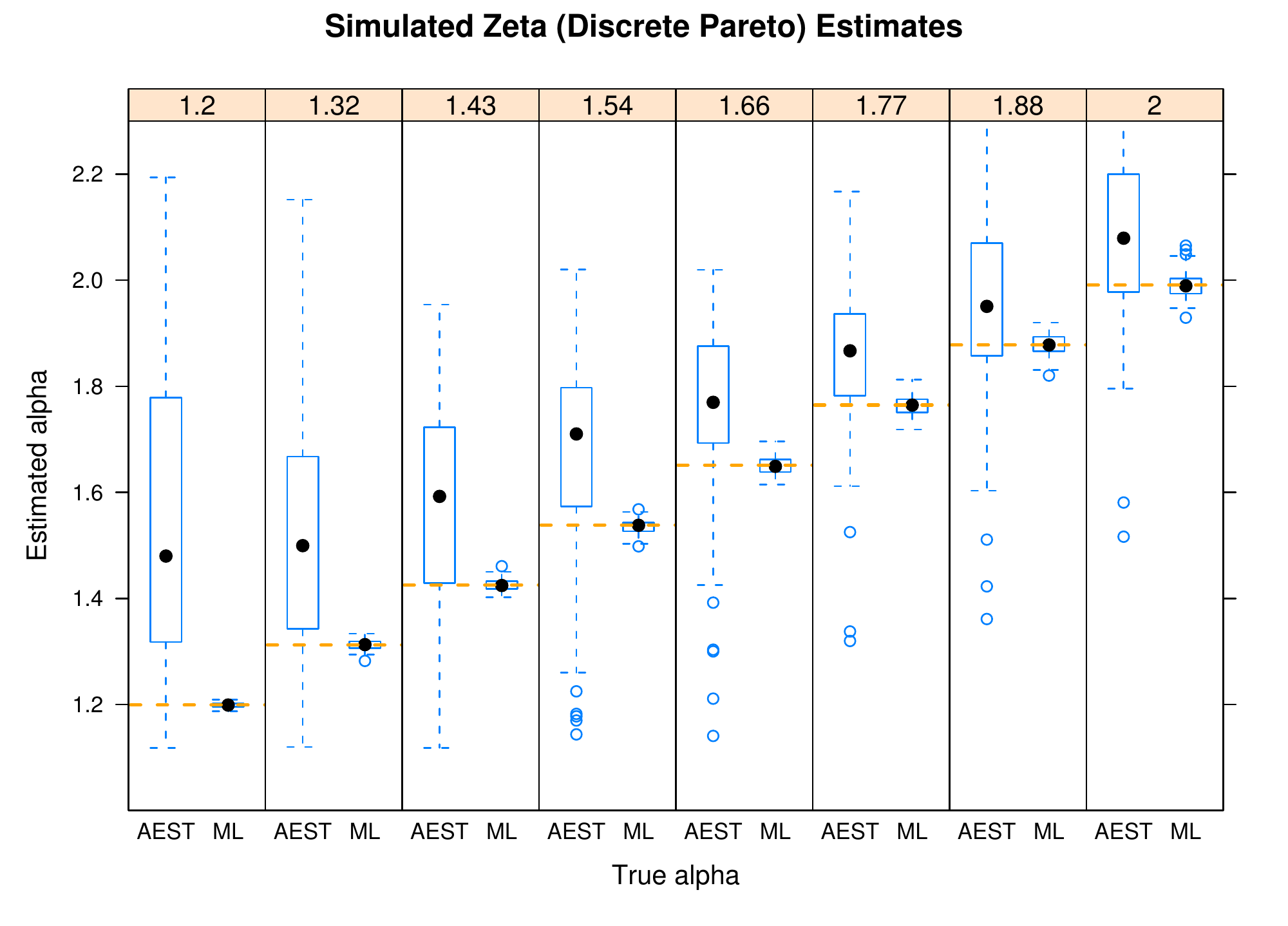}
\caption{When tested on synthetic Pareto-tailed data over $1< \alpha \leq 2$
an EP mixture model estimator performs significantly better than the well-known AEST method.}
\label{fig:compare_synthetic}
\end{figure}

Fig.~\ref{fig:compare_synthetic} compares the EP mixture model estimates of $\alpha$ on EP sample data over eight test values of $\alpha$ as indicated along the top of the plot. For each test value, we ran 100 test cases of 10000 samples each, with $m_1 = 0.5$ and $\lambda$ values chosen from the interval $[0.1.. 0.3]$. The box-and-whiskers plots show the distributions of  the 100 estimated $\alpha$ values by the two different methods, as compared to the true values shown by the dotted horizontal line.

The $\hat{\alpha}$ values box-and-whiskers obtained via the  {\bf aest} test are paired with the MLE plots on the same data in each panel of the figure.
We see that the range of $\hat{\alpha}$'s in the columns subtitled ``ML'' for the mixture model estimates, is almost always within a few percent of the true value.
Interestingly we see that the  {\bf aest}  $\hat{\alpha}$'s have a higher variance and tend to be biased.   In some sense, this is not surprising as the authors published results~\cite{Crovella99estimatingthe} acknowledge similar estimate variances when the underlying distribution is Pareto. Similarly the bias may well be due to the exponential component bleeding into the tail estimate, due to the method using a larger range of the sample, at the expense of less purity.
The {\bf aest} estimator has several fundamental
limitations. The Hill
estimator only uses the tail portion of the data, whereas the
MLE mixture model uses all the data. Hence, it throws away information
and cannot achieve the same efficiency as the MLE.
This validates that our MLE optimization algorithm for the EP mixture model converges to an accurate value when run on simulated EP data.

\subsection{Correct Model Selection}

Next we confirm that model selection by pair-wise comparison of BIC scores
does indeed select the right model, despite the model set consisting of nested models. Since the EEP model subsumes the other two, the model with more parameters will always fit better, so the correct choice is driven by the penalty dues to the BIC parameter count  term.

We ran 100 test cases over a range of sample sizes from 500 to 20,000 points, in the style of an empirical ``design of experiments'' to find what sample sizes were necessary to show adequate model selection results. We ran 3 pair-wise comparisons: EP vs. P on EP data, EEP vs EP on EP data and EEP vs EP on EEP data. In a fourth comparison, EP vs. P on P data, the informal results were so strong that it didn't merit a formal run.

In Table 2, we summarize the ability of our model selection method to distinguish the 3 hypotheses. For each test, we state the number of samples and the Bayes Factor level so achieved, using the conventions {\em substantial, strong, or decisive} in Table ~\ref{table:BF}.  For the first two tests we list sample sizes for two levels.

The more complicated the model comparison, the larger the sample required for the same strength of differentiation.  In short, the EP model can be selected ``substantially'' with around traces of no less than 1000 samples.  The EEP model requires about 10 times the sample to be selected at the same level.
This is reason to believe that requiring samples on the order of a few thousand (or at most 10,000) is a fairly light requirement for this class of models in our domain.

In practice, the next section reveals that the actual Bayes Factors computed on the data have values ranging in the hundreds, with sample sizes in the thousands and tens of thousands---clearly at the ``decisive'' level, and orders of magnitude larger than seen in these validation tests!  
Also, the $m_i$ mixture fractions mimic the model selection rules closely. This is to be expected, since $m_i$  represents the probability of the component $i$ in the sample. When its estimated value approaches zero, it is equivalent to selecting a model lacking that component. Were it not for the cost in sample size and computation one could always just estimate the most inclusive model and select the final model by eliminating components whose mixing fractions approach zero.

\begin{table}
\begin{center}
\begin{tabular}{|c|c|c|c|} \hline
Truth &  Model Choice: &  Min Number & $log_{10}$ BF  \\
      &                &  Samples    &  strength \\
    \hline
EP  & EP vs. P & 1000 &  substantial  \\
     &         & 5000 & decisive   \\
    \hline
EP  & EEP vs. EP  & 1000 & substantial \\
    &             & 10,000 & strong \\
    \hline
EEP  & EEP vs. EP & 9000 &  substantial  \\
    \hline
\end{tabular}
\label{tab:model_selection}
\caption{Sample sizes and the strength of comparison they achieve with simulated data, for different model comparisons.}
\end{center}
\end{table}

%

\section{Results}
\label{sec:results}

We now use our methodology to select the best model for each of our 270 users. We make some observations about our users based on the selected models and model parameters.

\subsection{Choice of Models}

\begin{figure}
    \includegraphics[width=0.9\columnwidth,height=1.8in]{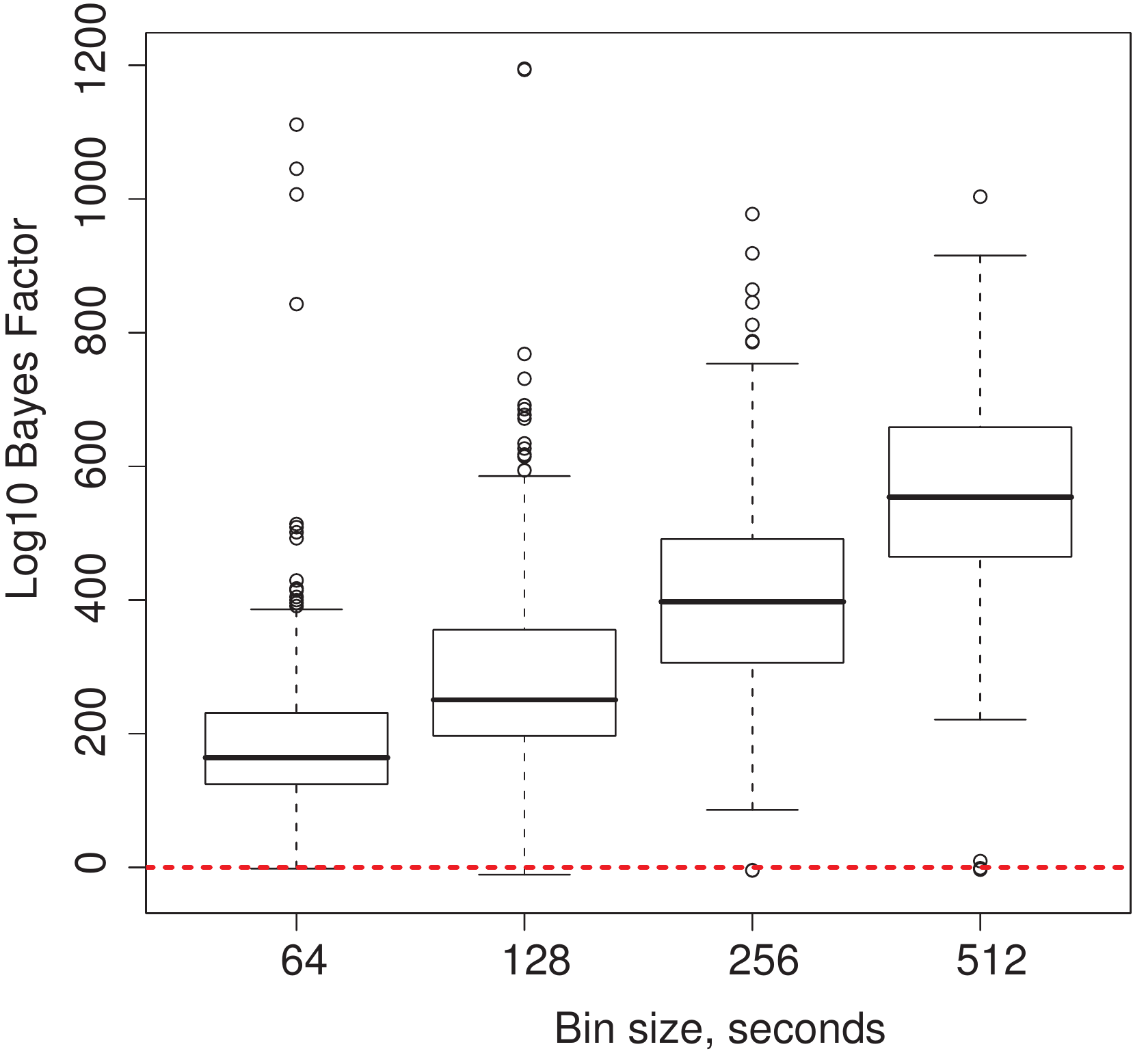}
    \caption{Boxplot of BIC comparison for Pareto vs. 2-component Mixture Model.}
    \label{fig:bic_pareto_2mixture}
\end{figure}

\begin{figure}
    \includegraphics[width=0.9\columnwidth,height=1.8in]{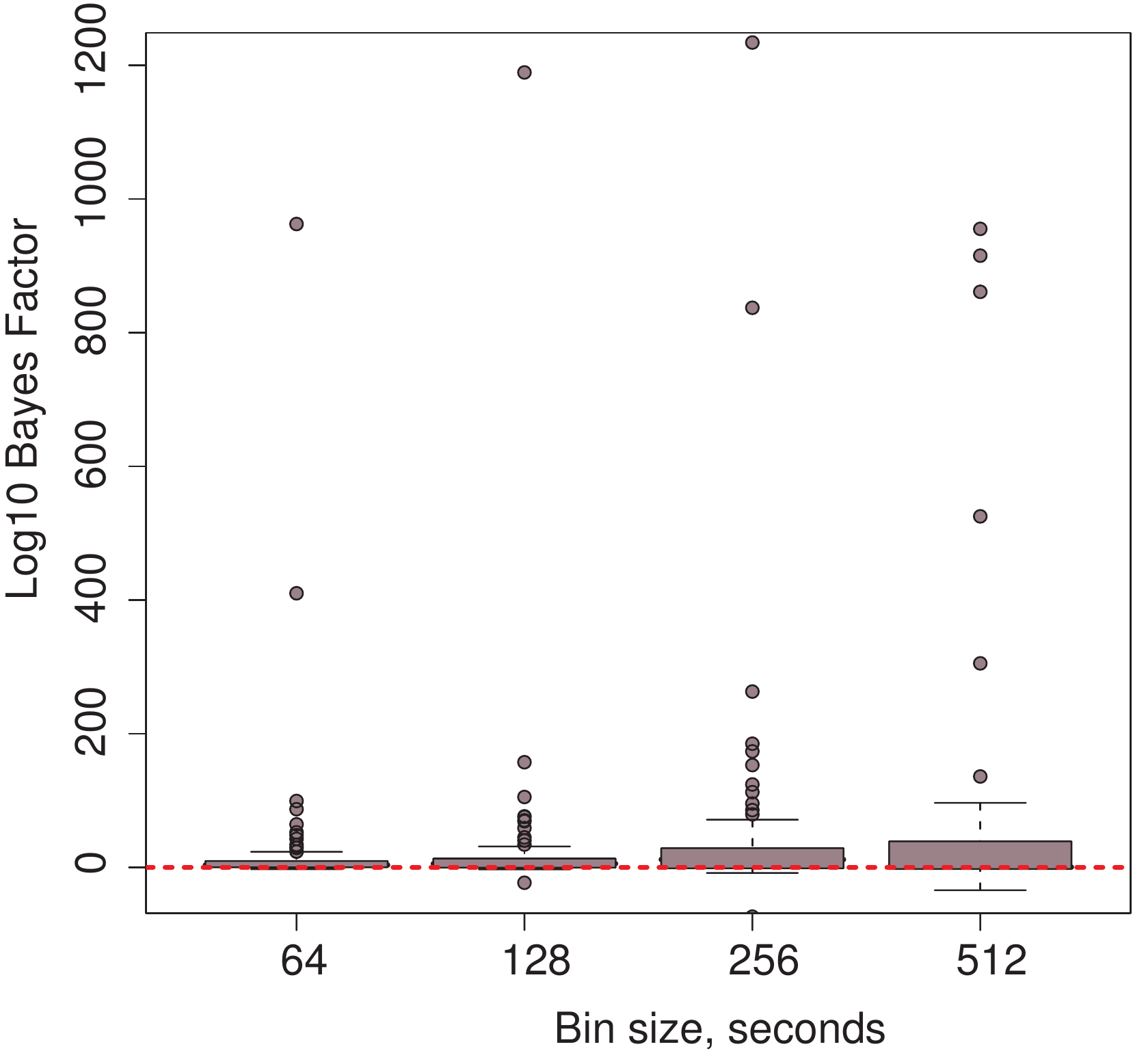}
    \caption{Comparison of EP and EEP models}
    \label{fig:bic_2mixture_3mixture}
\end{figure}

In Fig.~\ref{fig:bic_pareto_2mixture} we plot the log of the Bayes Factor (or difference in BICs) of the P and EP models. The x-axis labels indicate the bin size that was used when the models were computed. For each bin size, we computed the $\log BF_{EP,P}$ for each user. For each bin size, we use box plots to show the distribution of the Bayes factors over all the users. We can see that for nearly all users we can select the two component EP model as `decisive', according to Table 2. There are a very small number of users whose whose $\log BF_{EP,P}$ was near zero. Recall that when two models are considered indistinguishable, the model with fewer parameters is selected. The methodology selects a Pareto-only model for roughly a dozen of our endhost machines.
Not only is the two component mixture model EP preferred for all the other users, but it is strongly preferred as evidenced by the high Bayes factor values. We observe a small trend here in that as the bin sizes increase, the log Bayes factor ratio gets larger. This empirical observation indicates that for larger bin sizes, the exponential component plays an increasingly dominant role. Although not shown here on graphs, we also observed that as the bin sizes increase, the median mixing fraction $m$ increases. This corroborates the observation that the exponential plays an increasing role for larger bin sizes.

Next we compare the EP and EEP models.  Fig.~\ref{fig:bic_2mixture_3mixture} plots the distribution (as a boxplot) of the Bayes factors over all users, for each of 4 bin sizes. Interestingly, we see that at bin sizes of 64 and 128, the Bayes factors are close to zero for the majority of the users. Since the two models are fairly indistinguishable here, we again select the model of lower complexity, namely EP for nearly all the users. (There are a few outliers that would elect EEP). At larger bin sizes, we do see some users for whom the EEP model is selected. Overall, our method assigns the EEP model to roughly 30\% of the users and the EP model to the remaining 70\%.

\begin{figure}
    \includegraphics[width=0.9\columnwidth,height=1.8in]{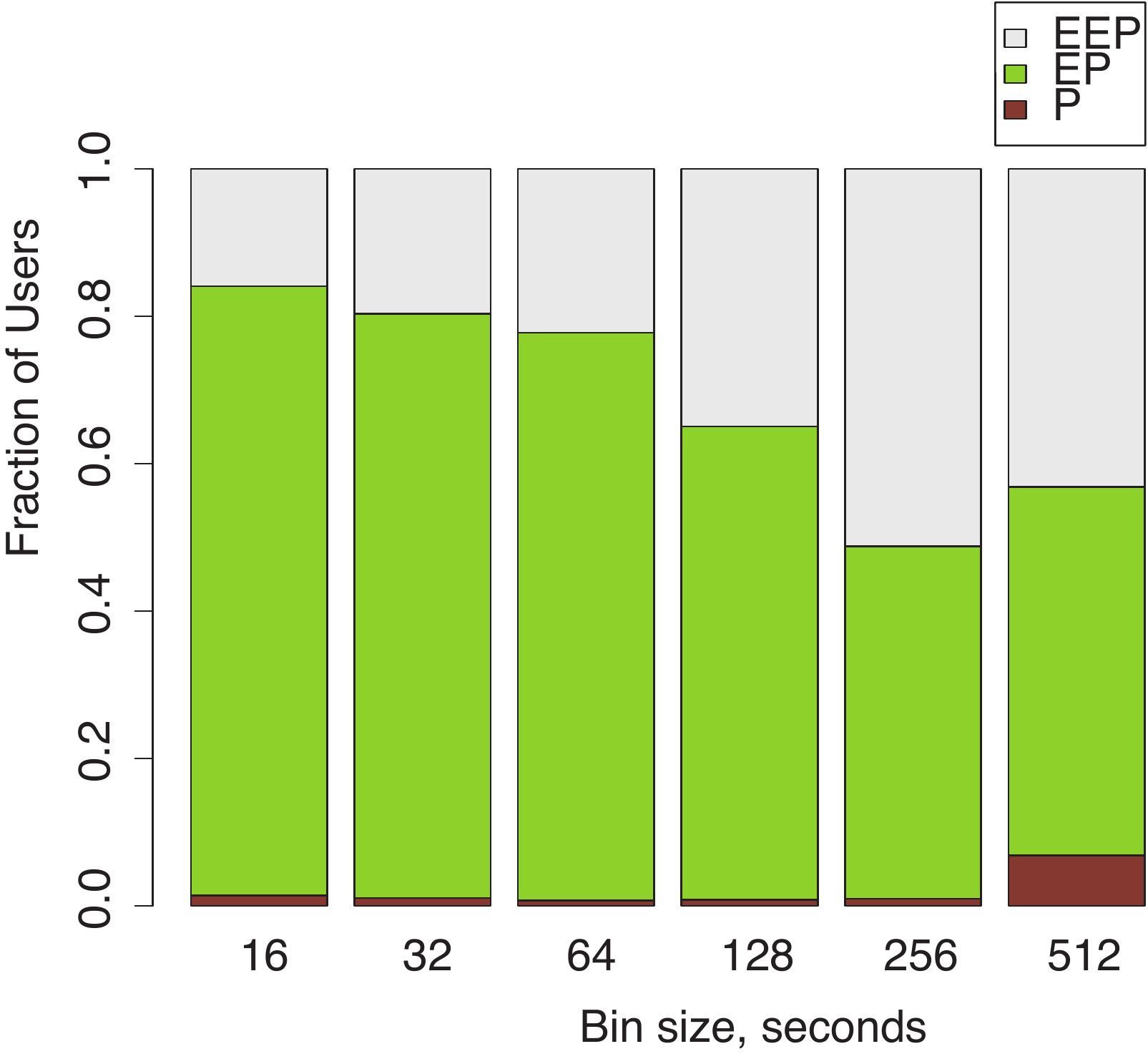}
    \caption{Choice of models based on Bayes Factors for different users. Each bar represents the same users with a different binning time window }
    \label{fig:model_selection_bar}
  \end{figure}

Fig.~\ref{fig:model_selection_bar} shows which model is selected by the methodology for all of our endhost machines.
Overall we see that only a handful of users are given the Pareto-only model, and between 15\%-40\% of  user machines are best modeled by  EEP (depending upon the bin size).
Overall, the EP model is selected for 50-85\% of the users, again depending upon the bin size. We conclude two things from this section. First, the flexibility we have built into our methodology is important and needed because the best model for one endhost is not necessarily the same for another endhost. Second, for the majority of the endhosts, the mixture model consisting of one exponential and one Pareto is clearly the preferred model.


\subsection{User Behavior}


As indicated in \S\ref{sec:related}, there is a growing interest
in understanding the range of variation of user behavior. We now look at
some model details to explore the range of parameters selected
across users, and the amount of mixing between the two model
components. We computed an EP model for all our users, and examined the
resulting $\alpha$ and $\lambda$ values.

We first observed that there is little correlation between $\alpha$ and
$\lambda$ values within each EP model.   This is reassuring, as it
indicates that the fitting process does not introduce dependencies
between the two component distributions, and that properties of one
distribution do not affect the other.

\begin{figure}
\includegraphics[height=2.5in,width=2.2in,angle=-90]{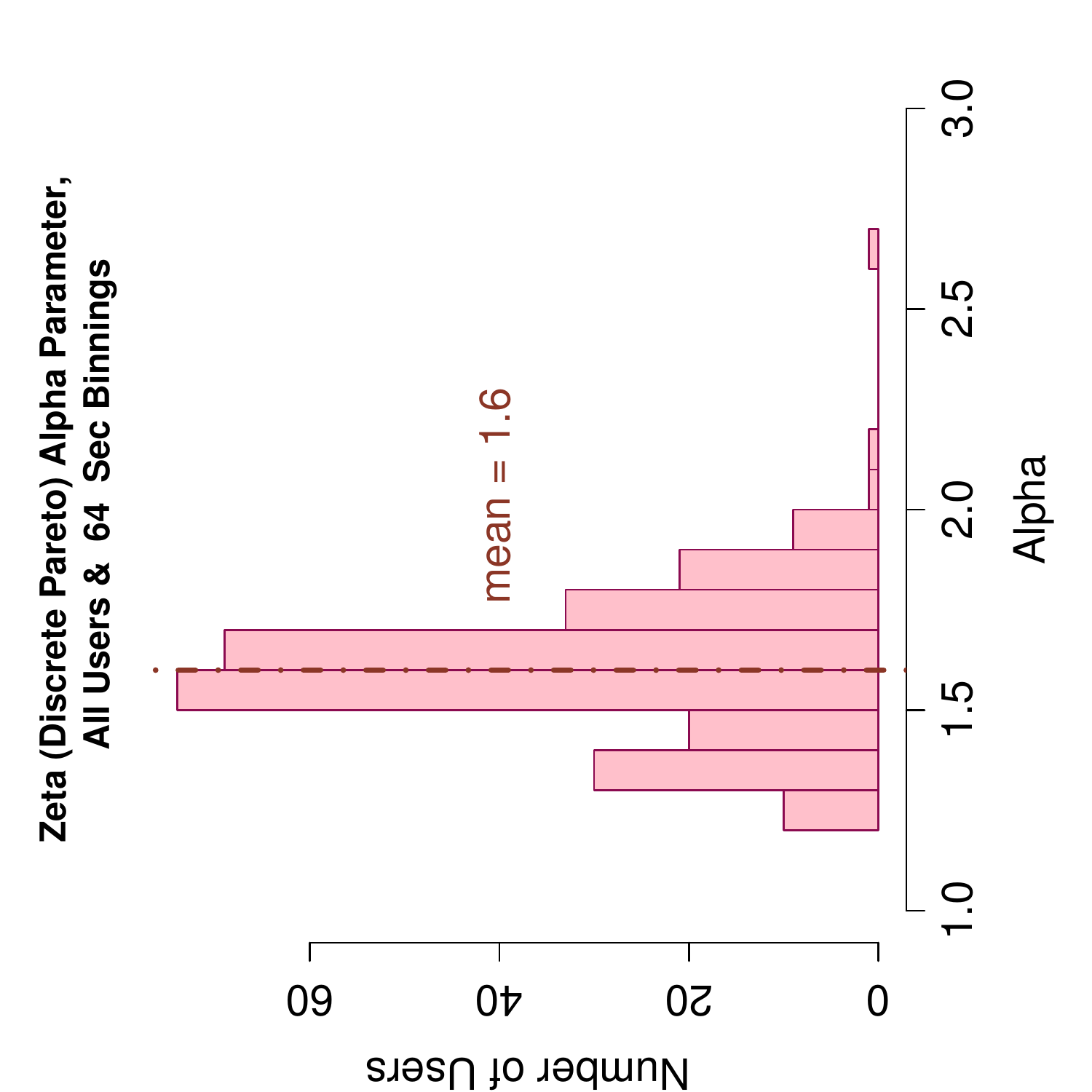}
\caption{Histogram of estimated $\alpha$ values across users.}
\label{fig:histogram_alpha_users}
\end{figure}

In Fig.~\ref{fig:histogram_alpha_users} we show the histograms of
$\alpha$ values over users for a bin size of 64 sec.
We see that the values of $\alpha$ range from 1.3 to 2.3 across the
users;  different users have very different properties in terms of the heaviness
of the tail of the distrbution.  Roughly 1/6 of our users have $\alpha <
1.5$ implying a fairly heavy tail, while most users have $\alpha$ values
around 1.6 or 1.7. It is interesting that we do have a small
number users (4) with $\alpha > 2$ indicating a finite second
moment.



\begin{figure}
\includegraphics[height=2.5in,width=2.2in,angle=-90]{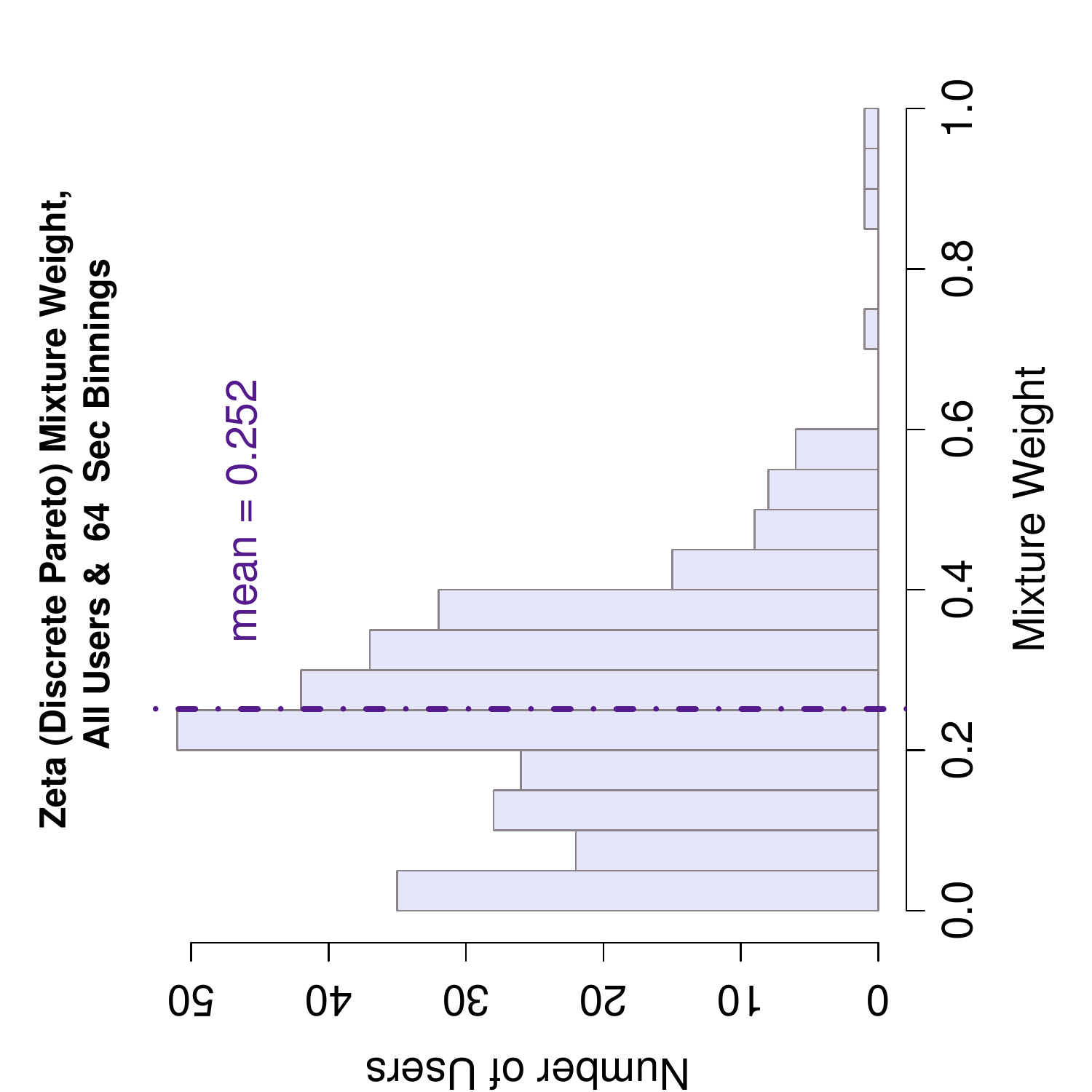}
\caption{Histogram of estimated $m_{\alpha}$ across users.}
\label{fig:histogram_mix_users}
\end{figure}

We now look more closely at how the users mix the two components of the
model. A value of $m$ close to 0 implies that the model is dominated by
the exponential distribution, (when $m=0$ there is no Pareto component
in the model). Similarly when $m$ is close to 1 the Pareto component
dominates the behavior of the model ($m=1$ indicates there is no
exponential component). The $m$ parameter is considered a {\em soft}
model selection factor because of its ability to indicate the strength
of each component of the distribution. The MLE estimates determine $m$
from the data, which is why it can be viewed as a soft model selection
factor. To see the range of $m$ values chosen across our users, we
provide a histogram of this mixing factor in
Fig.~\ref{fig:histogram_mix_users}. The frequency on the y-axis denotes
the number of users whose $m$ parameter is that indicated on the
x-axis. Only 3 users picked an $m$ very close to 1,
indicating that the pure Pareto model suites practically none of our
users---in agreement with the Bayes Factors conclusions. Most of the
users have an $m$ parameter less than 0.4, and roughly half of our users
had $m < 0.25$ indicating the dominance of the exponential component in
the model. The $m$ values are fairly well spread across the range 0 to
0.5 (roughly). We can also interpret this range of $m$ as a indication
of user diversity, in that their mixing fractions differ substantially.

\section{Traffic Composition}
\label{sec:applns}

The traffic models described are high level models that are
agnostic as to the particular kinds of applications or services
present in the traffic. There are several interesting questions that can be 
asked about the underlying generative processes that underlie 
the traffic. Are there particular applications and ports that tend to
occur more often in the exponential component of the distribution, or
the pareto component? Are there particular types of traffic that are
generated by human interaction, or by background processes on a
host? While a detailed analysis of such questions is outside the
scope of this paper, we present some initial findings towards these questions 
here.

We used our traces to see which applications are being used during
each of the two behavioral regions, 'exponential' and 'tail'. We can
soft-cluster the bins in each user trace (independently), as belonging
to the 'exponential' or 'tail' region of the model by comparing the
connection counts against our threshold that marks the start of the
tail. This clustering (or labeling) indicates which component of the
model is dominant in that window of time.   Using our keyboard and mouse click
data to associate with each bin a flag that
indicates if the user was active or idle in this time window. We
use a simple and conservative heuristic: the user is considered idle in a
time window if there was {\em no} recorded user activity in the
window, and active otherwise.

\begin{figure*}[ht]
\begin{minipage}[b]{0.5\linewidth}
\centering
\includegraphics[height=1.8in,width=\textwidth]{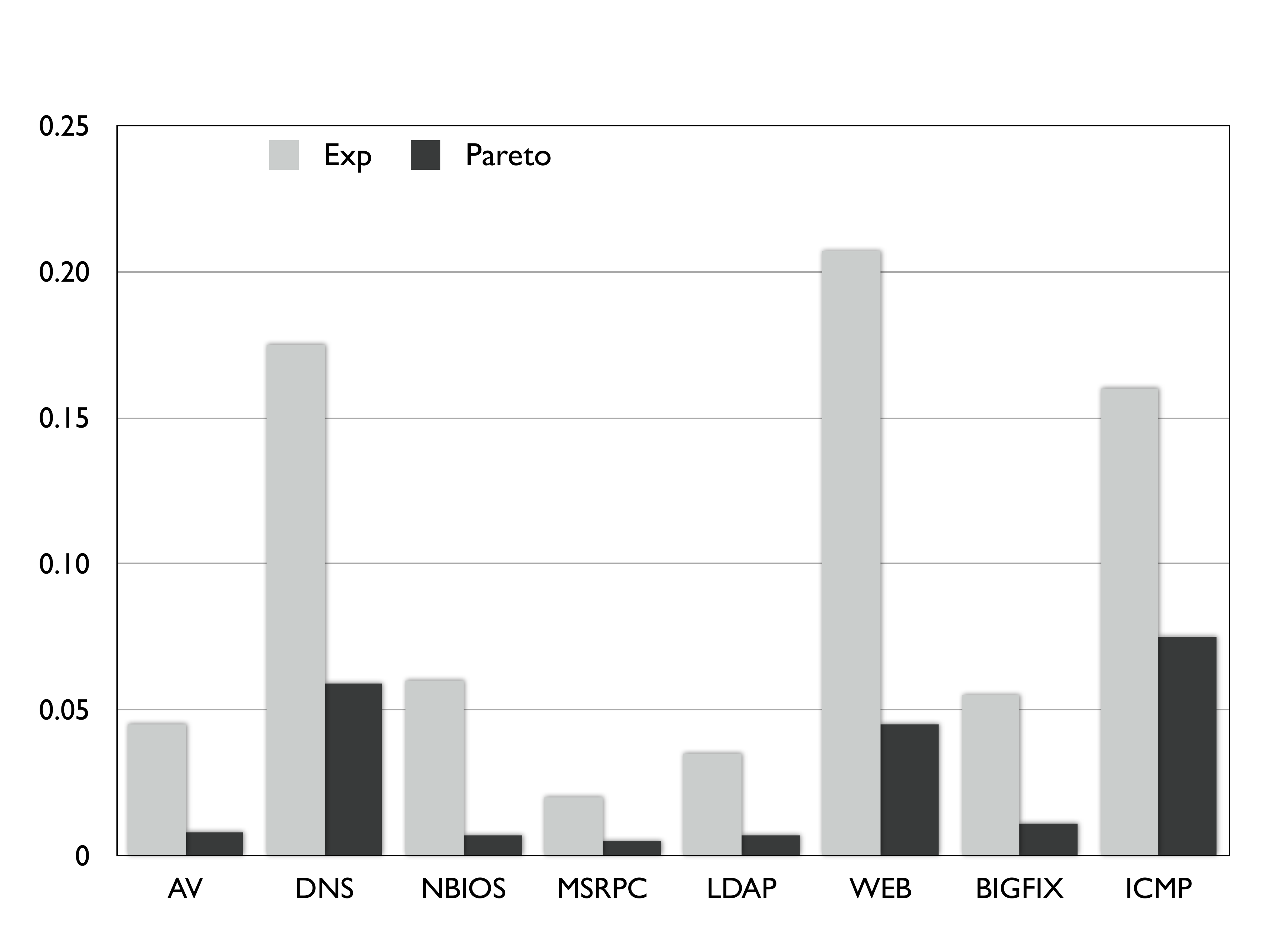}
\caption{Flow counts across bins marked 'exp' and 'pareto'}
\label{fig:e_vs_p}
\end{minipage}
\hspace{0.3cm}
\begin{minipage}[b]{0.5\linewidth}
\centering
\includegraphics[height=1.9in,width=\textwidth]{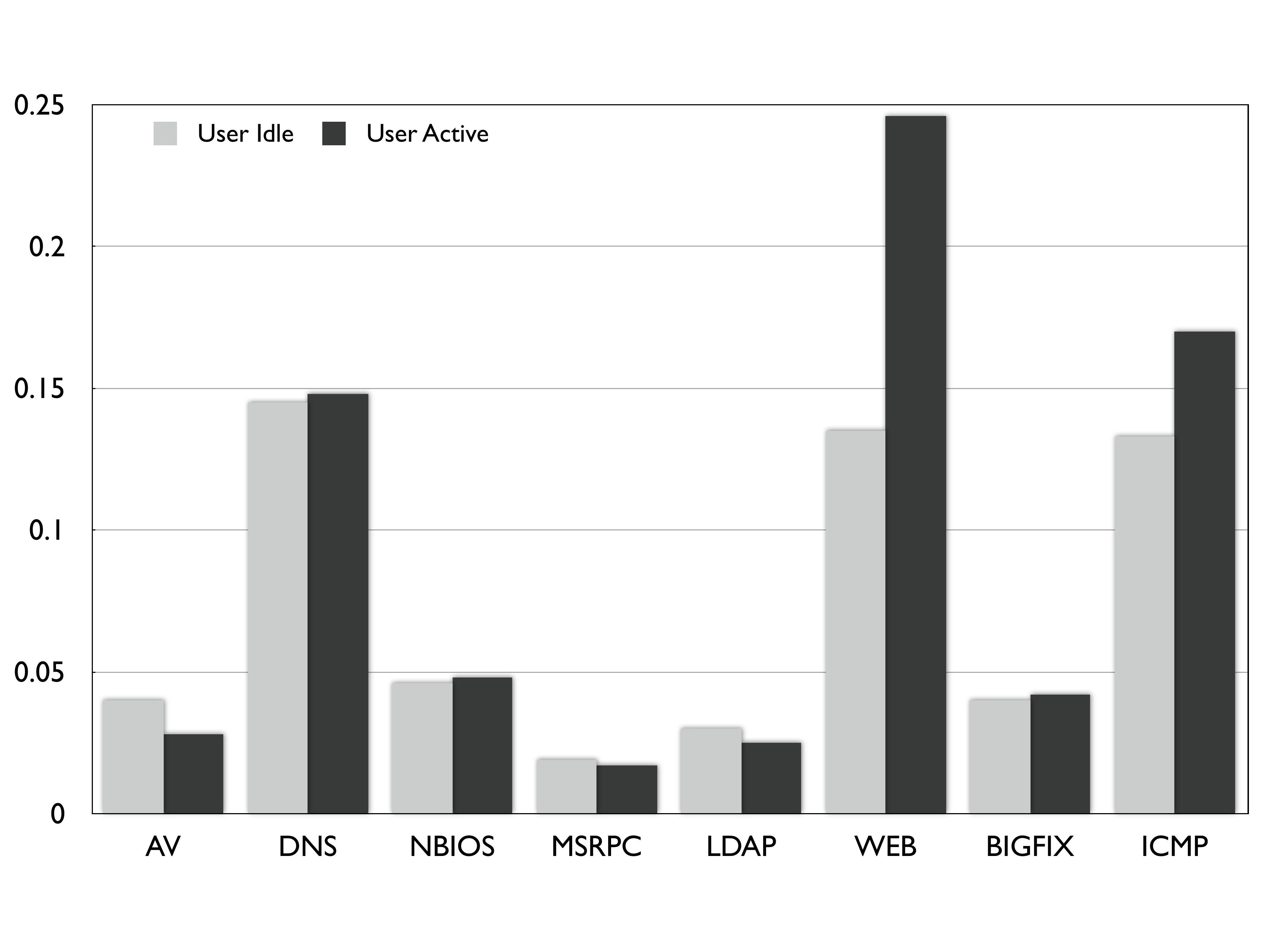}
\vspace{-2em}
\caption{Flow counts in bins where user was idle/active}
\label{fig:idle_vs_active}
\end{minipage}
\end{figure*}

We extracted the top 24 ports ranked by total count
across all the users and further semantically grouped them into a
smaller set of 9 traffic categories of interest. For instance, tcp traffic on ports 80, 443 and 8080 was
grouped into a ``web traffic'' category; we noticed dns traffic on
both tcp and udp, which was combined into a single ``dns traffic''
category. Fig.~\ref{fig:e_vs_p} plots the flow counts for each of these 9 traffic categories
 observed in bins falling in the exponential (or pareto)
part of the mixture model. The counts are normalized by dividing the
counts by the total flows observed in the exponential (or pareto)
bins, respectively. 
From the figure, we see that traffic in the pareto tails is dominated by four traffic categories, DNS, Web, ICMP and Bixfix.
Bixfix is an enterprise application that automatically manages software updates. Large ICMP bursts in our
enterprise are known to occur due to activities that scan multiple servers to find the closest one for a download.
The behavior during so called "exponential" bins (windows of time) appears
to be driven by all 8 categories shown, with Web, DNS and ICMP being the primary drivers.
One may postulate that the Web and DNS traffic
is primarily human-triggered activity.  ICMP is present to a roughly equal degree
in both the exponential body and the pareto tail; the implication
being that the ICMP ``bursts'' vary a great deal in size.


Fig.~\ref{fig:idle_vs_active} plots the flow counts for a particular
traffic category as it is observed in bins where the user is idle and
when the user is active. Again, the counts were normalized by the total flow
counts in each class. In this breakdown, we see that
most of the traffic categories examined are present in equal measure
whether the user is idle or active. On the one hand, the existence of a fair amount of heavy-tailed traffic
during user-idle periods is somewhat surprising because it opposes findings from other heavy-tailed research studies
claiming that user behavior is a cause of heavy-tailed traffic. 
On the other hand, it makes sense when you consider modern day practices for configuring enterprise clients.
Such clients come pre-configured with security, monitoring and management software, which run autonomously and generate traffic that does not depend on user presence. We see that web traffic is the only category that differs substantially between user-active and user-idle time periods. The web traffic during user-idle periods may reflect web content that is refreshed aggressively, and also asynchronous (eg. AJAX) style applications. 


While the results presented here are extremely preliminary, there is
evidence that points to specific applications (and traffic types)
contributing more to one part of the mixture model distribution. We plan to follow this direction in our future efforts.

\section{Conclusion}
\label{sec:conc}

In this paper we set out to model flow traffic as generated by endhost machines such as enterprise employee laptops.
We employ mixture models based on a convex combination of component distributions with both heavy
and light-tails. We approach the modeling problem as a model selection problem rather than a goodness-of-fit test. Our methodology selects the best model for an endhost by considering a family of 3 models
and doing pairwise comparisons to pick the best one. We employ the Bayes factor, based on the Bayesian Information Criteria (BIC), for these comparisons. To the best of our knowledge, this is the first paper to study heavy tails of data collected directly on endhosts, and is the first to employ a model selection approach.

We apply our methodology to data collected from 270 enterprise users, and find that for the vast majority
of users, the methodology selects the EP model. Although there are some users best modeled by EEP, and few by P. This shows the importance of a method that users a family of distributions and does
not presuppose a single distribution model for flow traffic.
We learn that our enterprise user population contains a great deal of diversity; not only do different users need different models, but some are heavy-tailed and others not. We observe a wide range of values for the tail slope and mixing fraction in our models. We take an initial glance deeper into the network traffic and see hints that a small number of protocols and applications may be responsible for the observed heavy tail behavior. We also see the presence of heavy-tailed traffic when users are idle indicating that the flows comes from machine-generated traffic (such as enterprise applications and chatty protocols).  In the future we plan to further explore the generative models behind the traffic patterns we observed herein. 

\bibliographystyle{acm}
\small
\bibliography{plaw}


\end{document}